\documentclass[preprint2]{aastex}
\usepackage{url}

\shorttitle{Neutral-ion drag in the chromosphere}
\shortauthors{Krasnoselskikh et al.}
\begin{document}

\title{Generation of electric currents in the \\
chromosphere via neutral-ion drag}
\author{V. Krasnoselskikh\altaffilmark{1}}

\affil{LPC2E, CNRS-University of Orl{\' e}ans\\
3A Avenue de la Recherche Scientifique\\
45071 Orl{\' e}ans CEDEX 2 FRANCE}

\author{G. Vekstein} 
\affil{School of Physics and Astronomy\\ The University of Manchester\\
Alan Turing Building, Manchester M13 9PL UK }

\author{H.~S. Hudson\altaffilmark{2}, S.~D. Bale, and W.~P. Abbett}

\affil{SSL, UC Berkeley, CA, USA 94720}

\begin{abstract}
We consider the generation of electric currents in the solar chromosphere where the ionization
level is typically low. We show that ambient electrons become magnetized even for weak magnetic
fields (30 G); that is, their gyrofrequency becomes larger than the collision frequency while ion motions
continue to be dominated by ion-neutral collisions.  Under such conditions, ions are dragged by
neutrals, and the magnetic field acts as if it is frozen-in to the dynamics of the neutral gas. However,
magnetized electrons drift under the action of the electric and magnetic fields induced in the reference
frame of ions moving with the neutral gas. We find that this relative motion of electrons and ions results
in the generation of quite intense electric currents. The dissipation of these currents leads to resistive
electron heating and efficient gas ionization. Ionization by electron-neutral impact does not alter the
dynamics of the heavy particles; thus, the gas turbulent motions continue even when the plasma
becomes fully ionized, and resistive dissipation continues to heat electrons and ions. This heating
process is so efficient that it can result in typical temperature increases with altitude as large
as 0.1-0.3 eV/km. We conclude that this process can play a major role in the heating of the
chromosphere and corona
\end{abstract}

\altaffiltext{1}{also SSL, UC Berkeley, CA, USA 94720}

\altaffiltext{2}{also University of Glasgow, UK G12 8QQ}

\section{Introduction}

The detailed physical mechanism of coronal heating is not yet well understood.  A number of
fundamental questions remain that challenge theoretical descriptions and the interpretation of
observational data (see, e.g. Klimchuk, 2006, or Walsh \& Ireland, 2003, for recent reviews).
In addition, the heating of the chromosphere requires much more energy, and this related process
is equally difficult to understand.
\nocite{2006SoPh..234...41K} 
\nocite{2003A&ARv..12....1W} 

\cite{1988ApJ...330..474P} proposed the idea that the solar corona could be
heated by the episodic dissipation of energy at many small-scale tangential
discontinuities arising spontaneously in the coronal magnetic field, as it
becomes braided and twisted by random photospheric footpoint motions. These
events -- sudden changes of the magnetic field topology -- hypothetically
result in plasma heating and the acceleration of non-thermal particles.
Parker invented a special name for these elementary energy release events, 
\textit{nanoflares}. The inspiration for this was the discovery of the hard
X-ray \textit{microflares} by \cite{1984ApJ...283..421L}; the energy of one
nanoflare was to be roughly 10$^{-9}$ times the energy of a major flare, and
thus orders of magnitude weaker than even the microflares. 
Parker's idea stimulated intensive searches  for any observational signatures 
of nanoflares
\citep{1994ApJ...422..381C,2008ApJ...689.1421S,2009A&A...499L...5V}
and their possible contribution
to the overall energy budget of the solar corona \citep{2006SoPh..234...41K}.

Microflares were first detected in hard X-rays in a balloon experiment by 
\cite{1984ApJ...283..421L}. The subsequent development of new
instrumentation allowed the multi-wave satellite and ground based
high-resolution observations of smaller-scale (about a thousand of kms or
even smaller) and lower energy phenomena. Soft X-ray imaging revealed
abundant microflares in active regions \citep{1994ApJ...422..906S}, and
RHESSI observations found that virtually all of a sample of some 25,000 hard
X-ray microflares occurred in active regions. \cite{1997ApJ...488..499K}
found flare-like brightenings in areas of the quiet Sun, and observations at
EUV wavelengths (e.g., the ``blinkers'' of \nocite{1997SoPh..175..467H} 
Harrison, 1997) reveal bursting activity
above the magnetic network borders. 
Similar phenomena that form small X-ray
jets at the limb were reported by Koutchmy et al. (1997). \nocite
{1997A&A...320L..33K} From these and other EUV observations \citep{%
1998A&A...336.1039B, 1998SoPh..182..349B}, if not the hard X-rays, we
conclude that Parker's idea of episodic heating of the apparently steady
quiet corona should not be discarded, even though no convincing evidence for
the required steepening \citep{1991SoPh..133..357H} of the energy
distribution function has yet been presented.

However, the idea of coronal heating via tangential discontinuities that arise spontaneously (nanoflares) does not address two important questions.  
Namely, where does the excess magnetic energy come from, and what was its original source?
Parker's analysis considered ideal MHD statistical equilibria that contain multiple discontinuities. 
Later \cite{2008ApJ...677.1348R} showed that a large-scale MHD energy source 
perturbed by slow motions on its boundary, supposed to be induced from the 
photosphere, results in the generation of a Poynting flux. 
This drives an anisotropic turbulent cascade dominated by magnetic energy. 
The result looks similar to Parker's tangling of magnetic field lines but the 
small-scale current sheets (which replace the tangential discontinuities) 
are continuously formed  and dissipated. 
In this modification of the initial scenario the current sheets are the result of the 
turbulent cascade. 
The initial energy reservoir in such a view is contained in large scale magnetic-field  
structures.

Here, we discuss another possibility: the direct generation of
relatively small-scale electric currents by neutral gas motions.
Clearly, a
huge energy reservoir exists in the form of turbulent motion of neutral gas
at and beneath the photosphere, supported by the underlying convection zone.
It is widely accepted that this energy can be partially transformed into the
excess magnetic energy in the chromosphere and corona. However, there is no
quantitative model that describes the physical mechanism of such energy
transfer. The development of a model that describes the energy transfer from
quasi-neutral gas motions to the magnetic field in a step-by-step manner
inevitably raises questions about the spatial and temporal scales at which
such a transfer can occur and about its location. Recent observations
provide strong indications that the energy reservoir is indeed the dynamic
turbulent photosphere, and that the energy transfer from the turbulent gas
motions to various kinds of trapped and transient magnetic field
oscillations takes place at the chromospheric level.

Analysis of the high resolution spatial ($\sim$150~km at the Sun) and
temporal (few sec) data obtained by the SOT (Solar Optical Telescope) aboard
Hinode (De~Pontieu et al., 2007) revealed that the chromosphere is dominated
by a multitude of thin ($\sim$200~km wide) dynamic, jetlike
``type~II'' spicules. 
\nocite{2007PASJ...59S.655D} 
They are ejected upwards
with characteristic velocities of 20-150 km/s and reach heights of
2000-10000~km before disappearing from the chromospheric passband (in this
case, the H~line of Ca~{\sc ii}). 
The type~II spicules have quite short lifetimes of
10-300~s (according to the authors, most of them last less than 100~s) and
many of them undergo substantial transverse displacements of the order of
500-1000~km. Moreover, the large-scale long-living spicules display
oscillatory motions in the direction perpendicular to their own axes. 
Since
the spicule structure can be taken to outline the direction of the magnetic
field, this led the authors to the conclution that the observed motions
indicate Alfv{\' e}nic oscillations.

Furthermore, the spatio-temporal variations of the chromosphere have always
revealed the greatest complexity, and prominences also consist of numerous
threadlike features with strong and mixed flows along these threads 
\citep{2005SoPh..226..239L}. Observations with SOT have provided an
exceedingly variable and dynamic picture of these flows and field structures %
\citep{2009ApJ...697..913O}. The SOT chromospheric data are in the H-line
(Ca \textsc{ii}), showing plasma at roughly 2~$\times$~10$^{4}$~K. The movie
presented by Okamoto et al (available on the Hinode web site %
\url{http://solarb.msfc.nasa.gov/}) shows ubiquitous continuous motions
along the prominence thread lines. The oscillatory motions observed might be
interpreted in terms of propagating or standing Alfv{\' e}n waves on the
magnetic field lines that presumably compose the prominence. The typical
transverse spatial scale of threads was found to be of the order of 600~km,
with a characteristic length of the order of several Mm. The characteristic
temporal scale was found to vary from 100 to several hundred seconds.

SOT observations near the limb led to the discovery of another small scale
dynamic phenomenon in the chromosphere: tiny chromospheric ``anemone jets,''
named for the similar X-ray features \citep{2007Sci...318.1591S}. They also
resemble larger-scale features well known in H$\alpha $ data and called
surges or sprays, which often occur near sunspots and in association with
flares or other transient activity (e.g. Rust, 1968). 
\nocite{1968IAUS...35...77R}
All of these observations can be considered as indications of the 
generation of small-scale perturbations and oscillations of the 
magnetic field in the chromosphere, where the degree of ionization is still 
relatively small.   

\section{The coupling problem}

In spite of greatly improved observational data, the physics of
chromospheric and coronal heating is not yet well understood. While the
underlying mechanism must be associated with the magnetic field, the details
of how efficiently the energy of convective motion is transformed into
magnetic energy in the solar atmosphere is an open question. The importance
of this process should not be understated. It is relevant not only to
coronal heating and the acceleration of the solar wind, but also to the
formation of the initial spectrum of wave turbulence introduced into the
solar wind.

We propose a mechanism to generate electric currents that can be viewed as 
Alfv{\' e}n or magnetosonic waves that result from strong ion-neutral drag. 
We assume that
photospheric motions are turbulent, and consist of both compressional and
rotational flows with energies exceeding that which is necessary to drive
the observed oscillations of suspended threads or jets in the chromosphere.
The question then becomes how these oscillations can be transported from the
photosphere to the chromosphere, and how the flow of neutrals is transferred
into the motion of charged particles and, ultimately, into the generation of
electric current.

There are two physical processes that are necessary to convert the energy of
the neutral gas motion into the magnetic field oscillations. Photospheric
motions of charged particles are dominated by frequent electron-neutral and
ion-neutral collisions, so that ions and electrons tend to follow the
neutral gas motion with the zero net electric current.

Since density falls off rapidly with height in the solar chromosphere, the
electron collision frequency with neutrals and ions decreases to the point
where it becomes smaller than the electron gyrofrequency. However, the
ion-neutral collision frequency at this height still remains substantially
larger than the ion gyrofrequency, since ion-neutral and electron-neutral
collision rates are not substantially different (see, e.g., De Pontieu et
al., 2001).
\nocite{2001ApJ...558..859D}
Therefore, the motions of
ions and electrons differ. Electrons tend to move along the magnetic field
lines and drift due to \textbf{E}~$\times $~\textbf{B} in the transverse
direction, while the ions continue to move together with the neutrals. 
This difference results in the generation of electric currents.

To view this situation from a different perspective, if ions move together
with neutrals (due to the strong drag) with some angle to the background
magnetic field, an inductive electric field \textbf{v}~$\times $~\textbf{B}
appears in the reference frame of the plasma. This induced electric field
inevitably generates electric currents that can be calculated using the
plasma conductivity tensor. When electrons and ions are both demagnetized,
this tensor reduces to the simple scalar conductivity, and the resulting
current is very small. When electrons become magnetized, the motions of
electrons and ions become decoupled, and the efficiency of the current
generation is substantially higher.

Thus, we conclude that there is an efficient chromospheric dynamo operating
in the layer between the level of electron magnetization and the height
where either the degree of ionization becomes high (comparable with unity)
or where the ions become magnetized. Forced oscillating currents are
generated most effectively when the inductive electric field has
characteristic frequencies and wavelengths close to the eigenmodes of the
system, i.e. the MHD waves in the magnetized weakly ionized plasma. At
higher altitudes, where ions are also magnetized, the motion of ions and
electrons can still differ. However, under these conditions, both components
will mainly move along the magnetic field lines and perform drift motions
across the field. Thus, the current is directed mainly along the magnetic
field lines, while in the intermediate region it can be generated in an
arbitrary direction.

In summary, understanding how currents are generated by the turbulent motions of neutral gas in the photosphere can be broken down into two steps.  
First, we must understand how the energy and vorticity of photospheric motions is transported upward to the level where electron and ion motions become decoupled. 
Second, we must derive a self-consistent system of equations that describe the inductive electric field and the resulting deformation of the background magnetic field.  
Once we obtain the electric field and estimates of current density, we can then determine the efficiency of electron heating due to collisions and can investigate the efficacy of other possible heating mechanisms.

\section{Atmospheric models}

\begin{figure}[h]
\centerline{\ \includegraphics[width=0.5\textwidth]{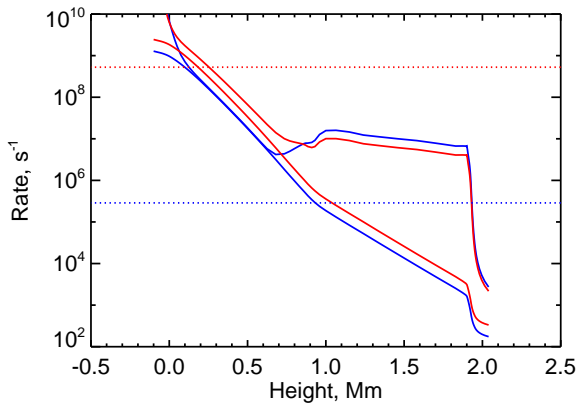} }
\caption{Collision frequencies in the quiet-Sun model of \protect\cite
{2009ApJ...707..482F}. Blue (red) curves represent proton (electron)
collision frequencies, with the lower lines showing neutral rates only 
\citep[e.g.][]{2001ApJ...558..859D}; the dotted horizontal lines show the
gyrofrequencies (ions and electrons) for an assumed field strength of 30~G. }
\label{fig:font_eH_pH}
\end{figure}

We appeal to standard semi-empirical model atmospheres for approximate
values of the physical parameters of the plasma in the photosphere and
chromosphere. \cite{2009ApJ...707..482F} provide a recent series of eight
such models representing features such as the quiet Sun, faculae, and
sunspots. Such models are based on macroscopic radiative transport theory
and are adjusted to recreate solar spectroscopic observations. 
They do not represent the dynamics, nor the plasma physics, since computer technology
thus far allows at best only an MHD approach to the physics %
\citep[e.g.,][]{2007AIPC..919...74S}. Figure~\ref{fig:font_eH_pH} compares
collision frequencies from the \cite{2009ApJ...707..482F} quiet-Sun model
(Model~1001) with the gyrofrequencies of electrons and ions. For this model we
assume a constant magnetic field of 30~G %
\citep[e.g.][]{2007ApJ...659L.177H,2008ApJ...672.1237L}.

\begin{figure}[h]
\centerline{\ \includegraphics[width=0.5\textwidth]{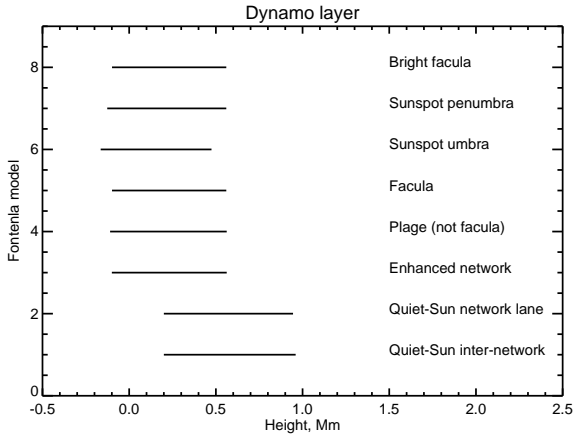} }
\caption{Chromospheric dynamo layers for the various solar features modeled
by \protect\cite{2009ApJ...707..482F}. The field strengths assumed are 30~G
for the quiet Sun (the lower two models), and 1500~G for the others. }
\label{fig:font_mag}
\end{figure}

The Fontenla models cover a range of physical features in the solar
atmosphere, and were originally intended for irradiance modeling. Here, we
use these models as a guide to set a range of parameters that are reasonably
consistent -- in this limited theoretical framework -- with solar structure.
Figure~\ref{fig:font_mag} shows the location of the chromospheric dynamo
layer for the full range of models. 
Electrons become magnetized, for all
models, essentially throughout the solar atmosphere. 
Thus the chromospheric
dynamo layer begins close to the photosphere and extends high into the
chromosphere. 
This is true for all of the Fontenla models, ranging from
sunspot umbra to bright facula, for which the transition-region pressures
range from 0.2~to about 2.0~dyne~cm$^{-2}$.
Note that in such models, substantial hydrogen ionization does not take
place below the transition region between chromospheric and coronal temperatures.

We emphasize that these models do not have any plasma physics in them as
such, and only serve as references at the order-of-magnitude scale. Indeed,
the mechanism discussed in this paper 
would point to areas in which models of the solar atmosphere can be greatly
improved.

\section{Energy transfer from the photosphere to the chromosphere}

We first consider the physics of mass, momentum, and energy transport
in the lower chromosphere. In order to formulate a
mathematical description of the problem we need to take into account
physical conditions between the photosphere and lower chromosphere, as
described above.

We begin with an isothermal atmosphere, i.e., a gas having a constant sound
speed and stratified by a uniform gravitational force acting in the negative
z-direction, for which 
\begin{equation}
\rho _{0}\left( z\right) =\rho _{ph}\exp \left( -z/H\right) .
\label{stat_eq}
\end{equation}
Here $\rho _{ph}$ is the density at the level of the photosphere
corresponding to $z=0$, 
and H~is the density scale height.
The gas motions are described by the
equation of continuity 
\[
\frac{\partial \rho }{\partial t}+\nabla \cdot {\{\left( \rho _{0}+\rho
\right) }\overrightarrow{v}\}=0, 
\]
where the density is presented as the sum of hydrostatic equilibrium density
and a perturbation (note that in hydrostatic equilibrium there are no flows).

By taking into account the equilibrium density dependence upon height this
can be re-written as 
\[
\frac{\partial }{\partial t}\left( \frac{\rho }{\rho _{0}}\right) +\frac{%
\partial v_{x}}{\partial x}+\frac{\partial v_{y}}{\partial y}+\left( \frac{%
\partial }{\partial z}-\frac{1}{H}\right) v_{z} 
\]
\begin{equation}
=\frac{1}{H}\frac{\rho v_{z}}{\rho _{0}}-\nabla \left( \frac{\rho }{\rho _{0}%
}\overrightarrow{v}\right) ,  \label{gas_density}
\end{equation}
where we have separated the linear and nonlinear parts and put the latter on
the right-hand side. The equation of motion 
\[
\left( \rho +\rho _{0}\right) (\frac{\partial \overrightarrow{v}}{\partial t}%
+(\overrightarrow{v}\nabla )\overrightarrow{v})=-\nabla P-\left( \rho +\rho
_{0}\right) G\widehat{z} 
\]
takes the form 
\[
\frac{\partial \overrightarrow{v}}{\partial t}+\nabla (\frac{P}{\rho _{0}})-%
\widehat{z}\frac{1}{H}(\frac{P}{\rho _{0}})+(\frac{\rho }{\rho _{0}})G%
\widehat{z} 
\]
\begin{equation}
=-(v\nabla )v-\frac{\rho }{\rho _{0}}(\frac{\partial v}{\partial t}+(v\nabla
)v).  \label{velocity}
\end{equation}
We shall also use the equation of state taken in the simplified form of the
equation for adiabatic motion, 
\[
\frac{dP}{dt}=\frac{\gamma P_{ph}}{\rho _{ph}}\frac{d}{dt}\left( \rho +\rho
_{0}\right) . 
\]
This is 
\[
\frac{dP}{dt}=C_{S}^{2}\frac{d}{dt}\left( \rho +\rho _{0}\right) . 
\]
It also can be re-written for perturbations as 
\[
\frac{\partial }{\partial t}(\frac{P}{\rho _{0}})-C_{S}^{2}\frac{\partial }{%
\partial t}(\frac{\rho }{\rho _{0}})+\frac{\gamma -1}{\gamma }\frac{C_{S}^{2}%
}{H}v_{z}= 
\]
\begin{equation}
C_{S}^{2}[(\overrightarrow{v}\nabla )\frac{\rho }{\rho _{0}}-\frac{v_{z}}{H}(%
\frac{\rho }{\rho _{0}})]-(\overrightarrow{v}\nabla )(\frac{P}{\rho _{0}})+%
\frac{v_{z}}{H}(\frac{P}{\rho _{0}}).  \label{pressure}
\end{equation}

Our objective here is the analysis of propagation of the photospheric
perturbations upwards into the chromosphere, and for this purpose we analyze
the characteristics of linear perturbations. To obtain the linear wave mode
dependencies, we determine the eigenmodes of the linearized system: 
\begin{equation}\label{gas_set1}
\frac{\partial }{\partial t}\frac{\rho }{\rho _{0}}+\nabla \cdot 
\overrightarrow{v}-\frac{1}{H}v_{z}=0\Rightarrow \frac{\partial }{\partial t}%
\frac{\rho }{\rho _{0}}=-\nabla \cdot \overrightarrow{v}+\frac{1}{H}v_{z},
\end{equation}

\begin{equation}
\frac{\partial \overrightarrow{v}}{\partial t}+\nabla (\frac{P}{\rho _{0}})-%
\widehat{z}\frac{1}{H}(\frac{P}{\rho _{0}})+(\frac{\rho }{\rho _{0}})G%
\widehat{z}=0,  \label{gas_set2}
\end{equation}
and 
\begin{equation}
\frac{\partial }{\partial t}(\frac{P}{\rho _{0}})+C_{S}^{2}\nabla \cdot 
\overrightarrow{v}-\frac{1}{\gamma }\frac{C_{S}^{2}}{H}v_{z}=0.
\label{gas_set3}
\end{equation}

We choose standard solutions of the form:

\[
P,\rho ,v\sim \exp i(\omega t-k_{x}x-k_{y}y-k_{z}z), 
\]
and analyze the dispersion relations that determine the dependencies of the
frequency $\omega $ upon components of the wave vector $\overrightarrow{k}$.
But our problem is formulated as a boundary-value problem where we assume
the perturbations and their time dependences to be prescribed at the
boundary $z=0$. 
We want to describe the evolution of perturbations with height~$z$.
In this case the very same equations should be used to find out
the dependence of the component $k_{z}$ of the $k$-vector upon the frequency 
$\omega $ and other components $k_{x}$, $k_{y}$. Substituting dependencies
defined above into the linearized equations one can obtain the dispersion
equation in the form 
\[
\lbrack \omega ^{2}-iKg]\{\omega ^{2}-C_{S}^{2}[(k^{2}+K^{2})+\frac{iK}{H}%
\frac{\gamma -1}{\gamma })]\}- 
\]
\[
iGC_{S}^{2}(k^{2}+K^{2})(K+\frac{i}{H}\frac{\gamma -1}{\gamma })=0, 
\]
where $C_{S}^{2}=\gamma GH$, $k^{2}=k_{x}^{2}+k_{y}^{2},K=k_{z}-\frac{i}{H}.$%
\[
k_{z}=\frac{i}{2H}\pm \sqrt{\frac{\omega ^{4}-\omega ^{2}C_{S}^{2}(k^{2}+%
\frac{1}{4H^{2}})+(\gamma -1)k^{2}g^{2}}{\omega ^{2}C_{S}^{2}}}. 
\]

This analysis demonstrates that a wide class of perturbations satisfying the
condition 
\[
\omega ^{4}-\omega ^{2}C_{S}^{2}(k^{2}+\frac{1}{4H^{2}})+(\gamma
-1)k^{2}G^{2}>0 
\]
increase with height, and the characteristic growth rate is 
\[
\Gamma = \mathrm{Im}(k_{z})=\frac{1}{2H}. 
\]

This phenomenon, the ``effective growth'' of the perturbations with altitude
in a hydrostatic gas equilibrium with exponentially decreasing density is
well known in the terrestrial atmosphere. It plays an important role in the
propagation of infrasonic perturbations induced by explosions in the
atmosphere (e.g. Klostermeyer, 1969a, 1969b) 
\nocite{Kloster1} \nocite
{Kloster2}. 
The same physical phenomena play important roles in the solar photosphere 
and chromosphere and have been the subject of many observational,
theoretical, and simulation studies.
At low frequencies acoustic-gravity waves (purely hydrodynamic) play
a role \citep[e.g.][]{2008ApJ...681L.125S,2009ASPC..415...95S},
and at high frequencies MHD modes \citep[e.g.,][]{1986JGR....91.4111H}.
The simulations \citep[e.g.,][]{1998ApJ...499..914S} can include 
essential physics such as optically-thick radiative transfer.

The results of the more realistic simulations highlight important features of neutral gas motion.  
For example, vorticity is primarily generated due to baroclinic forces, and tends to concentrate in tube-like structures whose widths are comparable to the numerical resolution.  
In addition, plasma heating depends on the formation of supersonic flows and shock-like structures.  
If shocks form systematically, they could well be an important factor in the physics of coronal heating. 
However according to \cite{1998ApJ...499..914S}, shocks are indeed observed at the edges of integranular lanes, but are a rare occurrence. 
At any one time, supersonic flow occurs in only 3-4\% of the surface area.
 
The role of acoustic gravity waves has also been intensively studied 
experimentally. 
Recently new experimental studies combining the SOT/NFI and SOT/SP 
instruments on Hinode and the Michelson Doppler Imager (MDI) on SOHO,
with the aid of 3D~computer simulations, were performed by 
\cite{2009ASPC..415...95S}. 
The authors came to the conclusion that the gravity waves are the
dominant phenomenon in the quiet middle/upper photosphere and that they
transport sufficiently more mechanical energy than the high frequency 
($>$5~mHz) acoustic waves. 
In addition they conclude that the acoustic-gravity wave flux is 3-5 times 
larger than the upper-limit estimate of \cite{2005Natur.435..919F}. 
These observations together
with the numerical models allow us to consider acoustic gravity waves
as one of the most important sources from which energy may be 
transformed to the magnetic field,
and then to heat via the action of electric currents. 

We note that that the absolute magnitude of the density
perturbation actually decreases with height, but it drops more slowly than
the background density. This results in the growth of the relative density
perturbation, written as ${\rho }/{\rho _{0}}$ in our notation. On the other
hand the \textit{velocity} perturbations really grow exponentially. An
important issue here is the characteristic vertical scale of this growth.
Taking the gas temperature to be 5000~K we find 
\[
H=\frac{k_{B}T}{MG}=140\ \mathrm{km}. 
\]
This linear analysis includes only the effect of the growth of the velocity
and relative density perturbations with the altitude, hence, the
perturbation scales do not vary with altitude. Of course, there are a wide
range of phenomena, such as vortical flows, that are not accounted for in
this formalism.

For example, the vortex radius can decrease with altitude. Indeed, assuming
conservation of angular momentum the increase of the velocity should result
in the shrinking of the transverse diameter of the vortex. 
This effect is
relevant for Rossby vortices in a multi-layer atmosphere and it is in
a good agreement with the observation of Stein and Nordlund (1998) that the
vorticity concentrates in small scale tubes.

\section{Magnetic perturbations produced by the turbulent motions of the
neutral gas}

Turbulent convective fluid motions at the photospheric level extend upwards
into the lower chromosphere. The latter is a weakly ionized plasma with a
typical temperature of 6000-7000~K, neutral hydrogen density $n_{H}$ $\sim $ 
$10^{12}-10^{14}$~cm$^{-3}$, and electron density $n_{e}$ $\sim $ $%
10^{10}-10^{11}$~cm$^{-3}$. Under these conditions the frequency of the
ion-neutral collisions is as high as $\nu _{in}\approx 10^{7}-10^{9}s^{-1}$
which, as will be confirmed below, produces quite a strong drag effect
resulting in the bulk velocity of ions, $V_{i}$ closely matching that of the
neutral gas, $V_{n}$. Then the magnitude of the generated electric current $%
\overrightarrow{j}=n_{e}e(\overrightarrow{V_{i}}-\overrightarrow{V_{e}}%
)\approx n_{e}e(\overrightarrow{V_{n}}-\overrightarrow{V_{e}})$ is
determined by the electron bulk velocity $V_{e}$. The latter is governed by
the equation of motion for the electrons 
\begin{equation}\label{el-mot}
m_{e}\frac{d\overrightarrow{V_{e}}}{dt}=-e\{\overrightarrow{E}+[%
\overrightarrow{V_{e}}\times \overrightarrow{B}]\}-(\nu _{en}+\nu
_{ei})m_{e}(\overrightarrow{V_{e}}-\overrightarrow{V_{n}}), 
\end{equation}
where $\nu _{en}\approx 5\times 10^{8}-10^{10}s^{-1}$ is the frequency of
electron-neutral collisions, the major source of the electron drag. 
Here $\nu_{ei}$ is electron-ion collision frequency.

Ions are supposed to be dragged by neutrals, thus the velocity of electrons can be expressed as
\[
\overrightarrow{V_{e}}=\overrightarrow{V_{n}}-\frac{\overrightarrow{j}}{en}. 
\]
The physical processes we describe occur in a parameter range
corresponding to the transition in electron motions from unmagnetized,
collision-dominated, to magnetized, when initially 
$|\omega _{Be}|\ <(\nu _{en}+$\emph{\ }$\nu _{ei}$), and then with the growth of
altitude $|\omega _{Be}|$  becomes larger than $(\nu _{en}+$%
\emph{\ }$\nu _{ei})$. 
We consider hereafter slow motions $\omega
<<(\nu _{en}+\nu _{ei}), |\omega _{Be}|$ similar to those we
described in the previous paragraph. 
In this case one can neglect the electron
inertia in the left-hand side of the equation \ref{el-mot}. 
It can then be simplified to obtain:

\begin{equation} \label{non_inertial}
-e\{\overrightarrow{E}+[\overrightarrow{V_{n}}\times \overrightarrow{B}]-%
\frac{1}{ne}[\overrightarrow{j}\times \overrightarrow{B}]\}+(\nu _{en}+\nu
_{ei})m_{e}\frac{\overrightarrow{j}}{ne}=0
\end{equation}

Taking the curl and replacing $\nabla \times B$ via 
$$
\frac{\partial \overrightarrow{B}}{\partial t}= - \nabla \times \overrightarrow{E} 
$$
one can obtain the final equation
\begin{equation}\label{magnetic_with_diff}
\frac{\partial \overrightarrow{B}}{\partial t} = \nabla \times [\overrightarrow{V_{n}}
\times \overrightarrow{B}] - \frac{1}{e\mu _{0}} \nabla \times \frac{1}{n}
[\nabla \times \overrightarrow{B}\times \overrightarrow{B}] 
$$
$$
- \frac{ (\nu_{en}+\nu _{ei}) m_e} {e^2\mu _0}
\nabla \times \frac{\nabla \times \overrightarrow{B}}{n},
\end{equation}
where $\overrightarrow{j}$ is replaced by 
\begin{equation}\label{Max}
\overrightarrow{j}=\frac{1}{\mu _{0}}\nabla \times \overrightarrow{B}. 
\end{equation}

Equation~\ref{magnetic_with_diff} together with the system of
equations \ref{gas_density}, \ref{velocity}, \ref{pressure} make up a closed
system describing an interaction of neutral gas perturbations with the ionized
electron component that can result in generation of electric currents. 
We neglect hereafter the influence of electron motions on neutral gas motions
by assuming that the degree of ionization is low. 

In what follows it is assumed that the chromospheric magnetic field
can be represented as
\[
\overrightarrow{B}=\overrightarrow{B_{0}}+\overrightarrow{b}, 
\]
where $\overrightarrow{B_{0}}$ is a background field, which for the
sake of simplicity is assumed here to be just a uniform field,
while ${\bf b}$~is a relatively small field deformation caused by a prescribed flow of 
neutral gas with velocity 
$\overrightarrow{V_{n}}$ and frequency $\omega $.

Then the linearized (with respect to ${\bf b}$) version of 
Eq.~\ref{magnetic_with_diff}
takes the form 
\begin{equation}\label{linearized}
\overrightarrow{b}+id_{e}^{2}\frac{\mid \omega _{Be}\mid }{\omega }(%
\overrightarrow{h}\cdot \nabla )\nabla \times \overrightarrow{b}%
-d_{e}^{2}(1+i\frac{\nu _{en}}{\omega })\nabla ^{2}\overrightarrow{b}=
\end{equation}
$$
i\frac{B_{0}}{\omega }\{(\overrightarrow{h}\cdot \nabla )\overrightarrow{%
V_{n}}-\overrightarrow{h}(\nabla \cdot \overrightarrow{V_{n}})\}. 
$$
where ${\bf h}$~ is the unit vector along the background magnetic field,
and $d_e = c/\omega_pe$ is the skin depth.

Let us now estimate the relative roles of the different terms in the left-hand side 
of Equation~\ref{linearized}.
For $n_{e}=10^{11}$~cm$^{-3}$ the electron
plasma frequency $\omega _{pe}=10^{10}$~s$^{-1}$, which
yields the $d_{e}\approx $~3~cm. 
The typical frequency of the neutral gas flows
under consideration is $\omega \approx (10^{-2}-10^{-3})$~s$^{-1}$ and their
length scale is $L\sim 10^{2}-10^{3}$~km. 
For $B_{0}\approx 100G$ the electron gyrofrequency 
$\omega _{Be}=2\times 10^{9}$~s$^{-1}$, 
while the characteristic electron-neutral collision frequency
$\nu_{en} \sim (10^8 - 10^9) \mathrm{sec}^{-1}$.
Therefore the ratio of the second and third terms compared to the first can be
considered small.
Thus Equation~\ref{linearized} can be reduced to
\begin{equation}
\overrightarrow{b}\approx i\frac{B_{0}}{\omega }\{(\overrightarrow{h}\cdot
\nabla )\overrightarrow{V_{n}}-\overrightarrow{h}(\nabla \cdot 
\overrightarrow{V_{n}})\},  \label{mag_approx}
\end{equation}
which means that in the lower chromosphere the magnetic field is effectively
frozen into the neutral-gas flow.

This describes the magnetic field generation due to the motions of
conductive fluid neglecting small scale effects scaling as the electron
inertial length, and assuming the motions to be slow: $\omega <<\Omega _{e}$.
Let us now verify that the assumption made above, namely that the bulk
velocity of the ions $\overrightarrow{V_{i}}$ is so close to that of the
neutrals $\overrightarrow{V_{n}}$ that the electric current can be written
as $\overrightarrow{j}=n_{e}e(\overrightarrow{V_{n}}-\overrightarrow{V_{e}})$%
. It can be easily found from the expression \ref{mag_approx} that 
\[
\overrightarrow{j}=\frac{1}{\mu _{0}}\nabla \times \overrightarrow{b}= 
\]
\[
i\frac{B_{0}}{\mu _{0}\omega }\nabla \times \{(\overrightarrow{h}\cdot
\nabla )\overrightarrow{V_{n}}-\overrightarrow{h}(\nabla \cdot 
\overrightarrow{V_{n}})\} 
\]
\[
\approx i\frac{B_{0}}{\mu _{0}\omega }(\overrightarrow{h}\cdot \nabla
)\nabla \times \overrightarrow{V_{n}}. 
\]

To evaluate the difference of the ion and neutral velocities let us consider
the equation of motion for ions: 
\begin{equation}
M_{i}\frac{d\overrightarrow{V_{i}}}{dt}=e\{\overrightarrow{E}+[%
\overrightarrow{V_{i}}\times \overrightarrow{B}]\}-M_{i}\nu _{in}(%
\overrightarrow{V_{i}}-\overrightarrow{V_{n}}).  \label{ions}
\end{equation}
As can be seen from Eq.~\ref{ions}, the sought-after velocity deviation $%
\overrightarrow{\Delta V_{i}}=(\overrightarrow{V_{i}}-\overrightarrow{V_{n}}%
) $ is such that the ion-neutral drag balances the other forces exerted upon
the ions. A simple estimate shows that the net electromagnetic force is the
dominant one there, and Eq.~\ref{ions} yields 
\[
M_{i}\nu _{in}\overrightarrow{\Delta V_{i}}\approx e\{\overrightarrow{E}+[%
\overrightarrow{V_{n}}\times \overrightarrow{B}]\} 
\]
\[
\sim \frac{1}{n}[\overrightarrow{j}\times \overrightarrow{B}]+\frac{m_{e}}{en%
}\nu _{en}\overrightarrow{j} 
\]
It follows then from Eq.~\ref{ions} that 
\begin{equation}
\overrightarrow{\Delta V_{i}}\approx \frac{e\mid \{\overrightarrow{E}+[%
\overrightarrow{V_{n}}\times \overrightarrow{B}]\}\mid }{M_{i}\nu _{in}}\sim 
\frac{\Omega _{i}}{\nu _{in}}\frac{\mid \overrightarrow{j}\mid }{ne}
\label{ion_pert}
\end{equation}
and 
\[
\overrightarrow{\Delta V_{i}}\sim \delta v_{e}\frac{\omega _{Bi}}{\nu _{in}}%
<\delta v_{e}, 
\]
where $\delta v_{e}$ is the velocity of current carrying electrons, and ions
are assumed to be demagnetized.

On the other hand, by using Eq.~\ref{mag_approx}, the electric current can
be estimated as 
\[
j\sim \frac{b}{\mu _{0}L}\sim \frac{B_{0}V_{n}}{\mu _{0}\omega HL}. 
\]
The velocity of the current -carrying electrons is 
\begin{equation}
\delta v_{e}\sim \frac{j}{ne}\sim \frac{eB_{0}V_{n}}{\mu _{0}m\omega HL}%
\frac{m}{ne^{2}}\sim \frac{\omega _{Be}}{\omega }\frac{d_{e}^{2}}{HL}%
V_{n}<V_{n}.  \label{veloc_comp}
\end{equation}

It is seen now from Eqs. \ref{ion_pert} and \ref{veloc_comp} that, indeed, $%
\delta V_{i}<<V_{n}$, and $\delta V_{i}<\delta v_{e}$.

A simple estimate of the currents generated at different altitudes and for
different values of the magnetic field can be obtained by assuming the
neutral gas motions to be rotational. In this case the characteristic
velocity in the vortex $V_{n}$ of characteristic spatial scale $L$ rotating
with the characteristic frequency $\omega $ can be evaluated as follows: 
\[
V_{n}\sim \omega L. 
\]
Thus, the characteristic magnetic field perturbation is 
\[
\delta B\sim B_{0}\frac{L}{H}, 
\]
and consequently the current density is 
\[
j\sim \frac{\delta B}{\mu _{0}L}\sim \frac{B_{0}}{\mu _{0}H}. 
\]

One can see that the magnetic field perturbations can become quite large;
the current densities for magnetic fields of the order of $100G$ can become
as great as 
\[
j\sim \frac{B_{0}}{\mu _{0}H}\sim 10^{-2}-10^{-1}\frac{A}{m^{2}}. 
\]

It is worth noting that this mechanism is dependent upon the velocity shear.
An interesting feature of this process is the very high efficiency
of the field generation around the boundaries of the neighbouring vortices,
where the characteristic shear of the velocity of the gas motions is large.
In this case the characteristic velocity shear is
\[
\frac{V_{n}}{L}>>\omega; 
\]
thus smaller-scale magnetic-field structures can be generated quite
efficiently and magnetic fields generated and current densities can become
tens or hundreds of times higher than the estimate above.

\section{Electron and ion resistive heating.}

To evaluate the electron heating efficiency one should find the induced
electric fields parallel and perpendicular to the magnetic field. This can
be done in a straightforward way by making use of the component of the equation
of motion for electrons along the magnetic field (Eq.~\ref{el-mot}). 
\[
m_{e}\frac{dV_{e\parallel }}{dt}=-eE_{\parallel }-\frac{\nu _{en}m_{e}}{ne}%
j_{\parallel }; 
\]
in the lowest-order approximation one should take into account that the
frequency $\omega $ of wave motions is much smaller than the electron
neutral collision requency. This leads to the resistivity along the magnetic
field to be classically collisional, where the dominant effect is due to
electron-neutral collisions 
\[
E_{\parallel }=-\frac{\nu _{en}m_{e}}{ne^{2}}j_{\parallel }=-\frac{\nu _{en}%
}{\varepsilon _{0}\omega _{p}^{2}}j_{\parallel }= 
\]
\[
-\frac{i\nu _{en}}{\mu _{0}\varepsilon _{0}\omega _{p}^{2}}\frac{B_{0}}{%
\omega }(\overrightarrow{h}\cdot \nabla\times \{(\overrightarrow{h}\cdot
\nabla )\overrightarrow{V_{n}}-\overrightarrow{h}(\nabla\cdot\overrightarrow{%
V_{n}})\}) 
\]
\[
\sim \nu _{en}\frac{d^{2}}{L}\delta B. 
\]

To evaluate the parallel electric field one should take into account that
the electron-neutral collision frequency varies with height from $\nu
_{en}\sim 5\times 10^{8}$ to $10^{10}$~s$^{-1}$ and the plasma frequency varies in
turn from $10^{10}$ to $10^{11}$. Taking the characteristic current
densities obtained above we find electric fields in the range 
\[
E_{\parallel }\approx (10^{-2}-10^{-1})\ \mathrm{V/m}, 
\]
with the electron heating written (Braginskii, 1965) as \nocite
{1965RvPP....1..205B} 
\begin{equation} \label{eq:brag}
\frac{3}{2}N_{e}\frac{d_{e}T_{e}}{dt}+p_{e}\nabla \cdot \overrightarrow{V_{e}%
}=-\nabla \cdot \overrightarrow{q_{e}}+Q_{e}.
\end{equation}
Here $p_{e}$ is the electron pressure, and the electron heating $Q_{e}$ is
determined by two ``friction forces,'' one due to the relative velocity
between electrons and ions/neutrals $(\overrightarrow{R_{ei}},%
\overrightarrow{R_{en}})$, and another due to the electron temperature
gradient $\overrightarrow{R_{T}}$: 
\[
Q_{e}=(\overrightarrow{R_{ei}},\overrightarrow{V_{e}}-\overrightarrow{V_{i}}%
)+(\overrightarrow{R_{en}},\overrightarrow{V_{e}}-\overrightarrow{V_{n}})+(%
\overrightarrow{R_{T}},\overrightarrow{V_{e}}-\overrightarrow{V_{n}}), 
\]
with
\[
\overrightarrow{R_{ei}}=-\nu _{ei}m_{e}N_{e}\{0.5(V_{e\parallel
}-V_{i\parallel })\overrightarrow{h}+(\overrightarrow{V_{e\perp }}-%
\overrightarrow{V_{i\perp }})\},
\]
\[
\overrightarrow{R_{en}}=-\nu _{en}m_{e}N_{e}\{0.5(V_{e\parallel
}-V_{n\parallel })\overrightarrow{h}+(\overrightarrow{V_{e\perp }}-%
\overrightarrow{V_{n\perp }})\},
\]
and
\[
\overrightarrow{R_{T}}=-0.7N_{e}(\overrightarrow{h},\nabla )T_{e}-\frac{3}{2}%
N_{e}\frac{\nu }{\Omega _{e}}[\overrightarrow{h},\nabla T_{e}]. 
\]
As a result 
\begin{equation}
Q_{e}=-\frac{(\nu _{ei}+\nu _{en})m_{e}N_{e}}{N_{e}^{2}e^{2}}%
\{0.5j_{\parallel }^{2}+j_{\perp }^{2}+
\end{equation}
\[
\frac{0.7N_{e}^{2}e}{\mu _{0}}(\overrightarrow{h}\cdot \nabla \times 
\overrightarrow{B})(\overrightarrow{h},\nabla )T_{e}\}; 
\]
here we neglect the diference between the ion and neutral velocities. The
last term in the equation of energy balance to be kept describes the heat
flux that is written as 
\[
\overrightarrow{q_{e}}=\overrightarrow{q_{ue}}+\overrightarrow{q_{Te}}, 
\]
\[
\overrightarrow{q_{ue}}=0.7N_{e}T_{e}(V_{e\parallel }-V_{n\parallel })+\frac{%
3}{2}N_{e}T_{e}\frac{\nu }{\Omega _{e}}[\overrightarrow{h},\overrightarrow{%
V_{e\perp }}-\overrightarrow{V_{i\perp }}], 
\]
and 
\[
\overrightarrow{q_{Te}}=-3.16\frac{N_{e}T_{e}}{m_{e}(\nu _{en}+\nu _{ei})}%
\nabla _{\parallel }T_{e}- 
\]
\[
4.66\frac{N_{e}T_{e}(\nu _{en}+\nu _{ei})}{m_{e}\Omega _{e}^{2}}\nabla
_{\perp }T_{e}-\frac{5}{2}\frac{N_{e}T_{e}}{m_{e}\Omega _{e}}[%
\overrightarrow{h},\nabla _{\perp }T_{e}]. 
\]
We shall begin by evaluating the energy supply provided by resistive
dissipation that is determined by first two terms in the right hand side of
the equation of energy balance. The volumetric heating power of the
electrons can be estimated as 
\[
\frac{3}{2}nk_{B}\frac{dT_{e}}{dt}\sim j_{\parallel }E_{\parallel }\sim 
\frac{\nu _{en}}{\mu _{0}}\frac{d^{2}}{L^{2}}\delta B^{2} 
\]
\[
\sim (10^{-4}\sim 10^{-3})\frac{W}{m^{3}}. 
\]

In a thermal equilibrium when the macroscopic flow evacuates the
power supplied by collisional resistive dissipation, one can estimate the
temperature variation with altitude as 
\[
\frac{dT_{e}}{dt}=V_{ez}\frac{\partial T_{e}}{\partial z},\frac{\partial
T_{e}}{\partial z}\sim \frac{j_{\parallel }E_{\parallel }}{\frac{3}{2}%
nk_{B}V_{ez}}, 
\]
where $V_{ez}$ is the characteristic macroscopic vertical velocity of
electron (plasma) motion that can be evaluated to be of the order of the
sound velocity. Then 
\[
\frac{\partial T_{e}}{\partial z}\sim \frac{j_{\parallel }E_{\parallel }}{%
\frac{3}{2}nk_{B}V_{S}}\sim (0.01-0.1) \ \mathrm{V/km}. 
\]

It is easy to see that the collisional heating of the electrons becomes
rather efficient. Another possible evaluation of the temperature variation
with altitude can be obtained by assuming a stationary static thermal
equilibrium. 
Under such conditions the power supply will be balanced by the divergence 
of the heat flux in Equation~\ref{eq:brag}, which can be estimated as 
\[
\nabla\cdot\overrightarrow{q_{e}}\approx \frac{\partial }{\partial z}\kappa
_{z}^{e}\frac{\partial T_{e}}{\partial z}, 
\]
where $\kappa _{z}$ is the electron thermal conductivity in the parallel
direction. For our conditions when electron-neutral conditions are dominant
in the parallel direction,
\[
\kappa _{z}^{e}\approx \gamma \frac{n_{e}T_{e}}{m_{e}\nu _{en}}, 
\]
where $\gamma $ is a coefficient of order unity (according to Braginskii,
1965) for electron-ion collisions; when ions are singly ionized it is equal
to $\sim $3.8. This leads to the following estimate of the characteristic
heat loss rate 
\[
\nabla\cdot\overrightarrow{q_{e}}\approx \frac{\partial }{\partial z}\gamma 
\frac{k_{B}n_{e}T_{e}}{m_{e}\nu _{en}}\frac{\partial k_{B}T_{e}}{\partial z}%
\sim \frac{k_{B}^{2}n_{e}T_{e}^{2}}{m_{e}\nu _{en}L_{T}^{2}}, 
\]
where $(T_{e}/L_{T})$ is the characteristic variation of temperature with
altitude. Under the physical conditions of the chromosphere this will lead
to 
\[
(\frac{T_{e}}{L_{T}})\sim (\frac{j_{\parallel }E_{\parallel }m_{e}\nu _{en}}{%
k_{B}^{2}n_{e}})^{1/2}\sim 0.1-0.3\ \mathrm{eV/km}. 
\]

Taking into account an exponential decrease of the plasma density one can
find that it can heat plasma to the hydrogen ionization energy (13.6~eV) on
a characteristic distance of the order of a few hundred~km. It should be
noted however that the thermal conductivity increases with the temperature
and this can result in some flattening of the temperature profile.

Another important physical effect is related to the ion heating due to the
perpendicular electric fields. It is known that the major component of the
current along the electric field perpendicular to the magnetic (the Pedersen
current) can be carried by ions. This can lead to the more efficient heating
of ions than of electrons. To evaluate the Pedersen current we must estimate
the ion velocity deviation from the velocity of neutrals in the direction of
the electric field, as was already done in the previous paragraph 
\[
\overrightarrow{\Delta V_{i}}\approx \frac{e\mid \{\overrightarrow{E}+[%
\overrightarrow{V_{n}}\times \overrightarrow{B}]\}\mid }{M_{i}\nu _{in}}\sim 
\]
\[
\frac{1}{M_{i}\nu _{in}n}[\overrightarrow{j}\times \overrightarrow{B}]\sim 
\frac{jB_{0}}{nM_{i}\nu _{in}}. 
\]
It follows then that 
\[
(n_e\overrightarrow{\Delta V_{i}}\cdot \{\overrightarrow{E}+[\overrightarrow{%
V_{n}}\times \overrightarrow{B}]\})\sim 
\]
\[
\frac{j^{2}B_{0}^{2}}{nM_{i}\nu _{in}}\sim \frac{M_{i}}{m_{e}}\frac{\Omega
_{i}^{2}}{\nu _{in}}\frac{d^{2}}{L^{2}}\delta B^{2}. 
\]
Ion heating due to the perpendicular electric field component can be
evaluated as 
\[
\frac{dT_{i}}{dt}\sim (\overrightarrow{j_{\perp i}}\cdot \{\overrightarrow{E}%
+[\overrightarrow{V_{n}}\times \overrightarrow{B}]\})\sim \frac{M_{i}}{m_{e}}%
\frac{\Omega _{i}^{2}}{\nu _{en}\nu _{in}}\frac{dT_{e}}{dt} 
\]

This estimate shows that the ion collisional heating can be several times
smaller than, or comparable to, the electron heating.

\section{Discussion}

The current generation mechanism that we propose here is strongly dependent
upon four basic parameters: the velocity of the turbulent gas motions, the
background magnetic field, the characteristic frequency of the spectrum of
turbulent motions of the gas, and the characteristic spatial scale of
velocity shear. 
The physical process of the generation of electric currents and magnetic
fields is actually nothing other than the 
well-known turbulent-dynamo mechanism operating as a result of the
the freezing of the magnetic field into the flow of the neutral gas.

The altitudes where this mechanism will operate are strongly dependent upon
the local magnetic field strength: the stronger the field, the lower the
altitude of electron magnetization. The efficiency of the electric current
generation is proportional to the magnitude of the velocity shear. The
characteristic spatial scales are proportional to the characteristic scales
of this velocity shear, i.e. the flow vorticity. It is interesting to note
that in this formalism, magnetic fields and currents can be generated by
both compressible and incompressible motions (see the first and second terms
in equation \ref{mag_approx}). Compressive motion tends to amplify or weaken
the background magnetic field, while the rotational component effectively
generates the perpendicular or helical components of the field. This means
that compressional motions tend to generate transverse currents, while
non-compressional motions can generate field aligned currents as well.

Since density tends to decrease exponentially with height in the
chromosphere, the characteristic velocity of chromospheric flows can
increase by a factor of ten or more over their photospheric values. 
This gives rise to an estimate of the characteristic magnetic field generated by
such motions readily becoming comparable to or even larger than the background
field.
Their relative magnitude is determined by the ratio of the characteristic
velocity of neutral motions at the altitude of electron magnetization to the
characteristic parameter $\omega H$. 
One can see that for characteristic
frequencies of the order of $10^{-2}-10^{-3}$~s$^{-1}$, and characteristic
heights of the order of 140~km, the field generation becomes quite efficient
already for the velocities of the order of ten-several tens of~km/s. The
helical magnetic field component can become larger than the background
field. A similar estimate shows that amplification/weakening of the
background magnetic field component can be evaluated to be of the order of 
\[
\frac{b}{B_{0}}\sim \frac{\delta n}{n}. 
\]
""Under such conditions, we show that the electron heating can become
rather efficient due to the collisional resistive dissipation of the
electric current. These processes take place in the regions where the
magnetic field vorticity increases.

It is worth noting that the efficiency of the electron and ion heating can
become substantially more important than that due to the collisional
resistive dissipation described above. 
The electric current and
magnetic field generation we have discussed takes place at the chromospheric
footpoints of flux tubes that may extend into the corona. The
direction of the generated field is determined by the characteristics of the
gas velocity shear and can lead to the formation of the magnetic field
configurations that can become unstable, or can form new structures by
interaction with neighbouring fields with different orientations. 
The increase of the perpendicular component of the magnetic field can result in
kink instability if this component satisfies the Kruskal-Shafranov condtion
\[
\frac{\delta B}{B}>\frac{l}{R}.
\]
Here $l$ is the characteristic scale of the cross section of the tube and $R$
is its characteristic length. Under such conditions the tube cannot keep
its configuration intact; it will become unstable and then different kinds
of perturbations will result in its reconfiguration. Other types of emerging
configurations with magnetic field inversions will result in local
reconnection. The detailed study of these effects lies beyond the scope of
our paper and will be carried out elsewhere.

The result of the heating would be a rapid increase
of the degree of ionization of the gas. 
It is worth remembering that our whole
description is valid only when the degree of ionization is low. 
This rapid ionization process can lead, according to most radiative-convective 
models \citep[e.g.,][]{1998ApJ...499..914S} to an efficient decrease of 
the radiative cooling rate of the gas. 
Thus the generation of electric currents results not only in the plasma
heating itself, but also causes a decrease of the cooling process
that will then strengthen the effect of radiative heating. 
We cannot make any quantitative estimate of these effects, but they may 
be adressed in future simulations.

The coupling of acoustic-type waves and MHD waves can also be
treated in the MHD approximation \citep[e.g.,][]{1982SoPh...75...35H}.
The major difference in our approach here  is that the neutral
collisions reduces the feedback effect on the motion of the gas
resulting from the newly generated magnetic field. 
This produces a more efficient transfer of energy from gas motions to 
current generation and magnetic field amplification.

It seems possible to incorporate certain aspects of this
formalism into a system of conservation equations similar to those used in
\cite{2007ApJ...665.1469A}. 
In that work, the resistive MHD equations were solved
numerically within a computational domain that includes both a turbulent
model convection zone and corona. We are currently updating this
system of equations to incorporate additional physics, and are
planning to perform a comparison between a standard resistive 
radiative-MHD model and our new results. 
We hope to report on the results of this study in the near future.

\section{Conclusion}

We have presented an analysis of the effect of electric current and
associated magnetic fields generation in a ``chromospheric dynamo layer''  
where electrons become magnetized while ions remain
collisionally coupled to the neutrals. 
We have shown that electric currents and
magnetic fields can be generated very efficiently due to turbulent motions
of the neutral gas. 
The efficiency of this physical process is proportional
to the characteristic velocity shear of the gas turbulent motions and inversely
proportional to their characteristic frequency. 
We also found that the magnetic fields thus generated can be comparable to 
or even larger than the background
magnetic field for motions having characteristic scales of the order of 
several hundred~km and characteristic time scales of the order of several minutes.
This can produce a substantial restructuring of magnetic
field configurations and an opportunity to create multiple sites of
reconnection. 
It also results in an efficient increase of collisional resistive heating of
electrons and ions, and hence the rapid ionization of the gas,
thus unstably altering its thermal equilibrium.

\acknowledgments

The authors are grateful to CNES for financial support in the frame of CNES
``Solar Orbiter Research grant'' and to SSL for the financial support of the
visit of V.~K. to the University of California at Berkeley. H.~H. thanks
NASA for support under NAG5-12878. V.~K. acknowledges very useful
discussions with M.~Ruderman, B.~De~Pontieu, O.~Podladchikova, and 
X.~Valliers.
The authors also thank the referee, whose comments helped in improving 
the article.

\bibliographystyle{apj}
\bibliography{krasno}

\end{document}